\documentclass[12pt,preprint]{aastex}

\usepackage{graphics}
\usepackage{color}
\usepackage{natbib}

\newcommand{\cmtwo}{cm$^{-2}$}  
\newcommand{\cmthree}{cm$^{-3}$}

\newcommand{\kms}{km\,s$^{-1}$}       %km/s
\newcommand{\es}{erg s$^{-1}$}                          %energy cgs
\newcommand{\ecs}{erg cm$^{-2}$ s$^{-1}$}

\newcommand{\um}{$\mu$m}                                 %micron
\newcommand{\lsun}{$L_{\odot}$}                          %solar and terr. units
\newcommand{\msun}{$M_{\odot}$}
\newcommand{\rsun}{$R_{\odot}$}
\newcommand{\mdot}{{\it \.{M}}}
\newcommand{\msunyr}{$M_{\odot} \, {\rm yr}^{-1}$}

\newcommand{\lapprox}{$\stackrel {<}{_{\sim}}$}
\newcommand{\about}{$\sim$}                       %approx
\newcommand{\powten}[1]{10$^{#1}$}
\newcommand{\halpha}{H$\alpha$}                   %H I recombination lines 

\newcommand{\oishort}{[O\,{\sc i}]\,63\,$\mu$m}

\newcommand{\av}{$A_{\rm V}$}                     %extinction

\newcommand{\asec}{$^{\prime \prime}$}
\newcommand{\adeg}{$^{\circ}$}

\newcommand{\asecdot}[2]{\mbox{#1$\stackrel {\prime \prime}{_{\bf \cdot}}$#2}}

\slugcomment{Date: \today}

\shorttitle{L\,1551 IRS\,5 binary star-jet system}
\shortauthors{Liseau, Fridlund \& Larsson}

\begin{document}

\title{Physics of Outflows: the Binary Protostar L\,1551 IRS\,5 and its Jets}

\author{Ren\'e Liseau\altaffilmark{1}}
\affil{Stockholm Observatory, AlbaNova University Center for Physics, Astronomy and Biotechnology,
SE-106 91 Stockholm, Sweden}
\email{rene@astro.su.se}
\and

\author{C. V. Malcolm Fridlund\altaffilmark{2}}
\affil{Astrophysics Division, ESTEC/ESA, P.O. Box 299, NL-2200AG Noordwijk, The Netherlands}
\email{malcolm.fridlund@esa.int}

\and

\author{Bengt Larsson\altaffilmark{1}}
\email{bem@astro.su.se}

\begin{abstract}
Recent observations of the deeply embedded L\,1551 IRS\,5 system permit the detailed examination 
of the properties of both the stellar binary and the binary jet. For the individual components
of the stellar binary, we determine their masses, mass accretion rates, effective temperatures
and luminosities. For the atomic wind/jet flow, we determine the mass loss rate, yielding
observationally determined values of the ratio of the mass loss to the mass accretion rate, $f$. 
For the X-ray emitting region in the northern jet, we have obtained the jet-velocity and derive the 
extinction and the densities on different spatial scales. Examining the observational
evidence within the framework of the x-wind theory leads us to conclude that these models 
are indeed potentially able to account for the observational data for this {\it deeply embedded} source.
\end{abstract}


\section{Introduction}

Since their discovery nearly three decades ago, the unexpected phenomenon of outflows 
in star forming regions has remained essentially unexplained. In particular, the processes
responsible for the acceleration and collimation of the flows present one of the major 
unsolved problems of modern astrophysics \citep{lada85,shu87,shu00,eis00,kp00}. Considerable progress 
has been made in the theoretical field, but observational results had generally been of 
too poor quality to make direct comparisons meaningful. In this paper, we examine
recent various observational results for the {\it deeply embedded} source L\,1551 IRS\,5, which,
when put together, finally permit the detailed comparison with theoretical models. 
Optically visible young stellar objects (lower mass loss and accretion rates) have previously 
been addressed by, e.g., \citet[and references therein]{shang02}.

Over the years, IRS\,5 has enjoyed a variable status of `the archetypical CO outflow source'
and that of `a pathological case' \citep[e.g.,~][]{padman97}. More recently, renewed
interest in this object has arisen, in part due to the recognition of its duplicity,
as binarity seems to be a very frequent phenomenon among powerful outflow drivers (M.\,Barsony, 
private communication). If protostellar binarity is indeed intimately related to the physics of
generating bipolar outflows, then one obviously wishes to understand the physics of
the binary itself. We devoted therefore quite some effort to arrive at the understanding
of the IRS\,5 system. Based on the theory of protostellar structure and evolution, we find a
solution capable of explaining observed characteristics of both IRS\,5 itself and its associated
jet flows, and which, when applied to x-wind models, successfully recovers the physical
properties of the jets. Thus, in this particular case, a direct dependence on source binarity
is not evidenced, as the theory of x-winds has been developed for single stellar objects.

The organisation of this paper is as follows: In Sect.\,2, we review the evidence based on recent 
observations in both the optical, infrared and radio spectral regimes. In Sect.\,3, we use published
models of protostellar structure and evolution to derive the physical properties of the individual 
binary components and apply x-wind theory to derive some parameters relevant in the present context.
In Sect.\,4, we discuss our results and, finally, in Sect.\,5, we briefly summarise our main conclusions.

\section{The Observational Evidence}

\subsection{Mass loss rates from the large scale atomic flow}

\citet{hollen85} proposed that the luminosity of the \oishort\ line can be used to estimate 
the wind mass loss rate. At the distance of 150\,pc and for the observed line luminosity with an 86\asec\ beam
\footnote{This refers to the flux measured by the ISO-LWS (Infrared Space Observatory Long Wavelength Spectrometer)
toward IRS\,5. In addition, another ten positions were observed toward the two CO outflow lobes.}
\citep{white00}, this method would yield a mass loss rate of $8 \times 10^{-7}$\,\msunyr. 
In contrast, based on H\,I 21\,cm observations of the redshifted gas with the VLA (Very Large Array, 
\about\,57\asec\ beam), \citet{giova00} determined the mass 
loss rate from IRS\,5 for the atomic wind as \mdot$_{\rm w,\,red}= 1 \times 10^{-6}$\,\msunyr.
Given the momentum symmetry of the red- and blueshifted flows (see their Table\,2), a mass loss rate of about
$2 \times 10^{-6}$\,\msunyr\ seems indicated, a factor of three higher than that based on the \oishort\ line.

The reason for this discrepancy is not entirely clear. One possibility could be
that the condition, in which the Hollenbach relation is thought to hold, is not met by 
the flow from IRS\,5 (viz. $\Phi_{\rm H}$ \lapprox\ $n_{\rm 0,\,5}\,v_{\rm s,\,100}$, 
where $\Phi_{\rm H}$ is the particle flux
in cm$^{-2}$\,s$^{-1}$ through the shock front, and when the pre-shock density $n_{\rm 0}$ is expressed 
in units of $10^5$\,\cmthree\ and the shock velocity $v_{\rm s}$ is in units of 100\,\kms).
To some degree this might be supported by the fact that IRS\,5 falls out completely, by more than one order 
of magnitude, of the empirical $L_{\rm forbidden\,line}-L_{\rm bol}$ relation found for a large number of jet-sources 
\citep{edwards93,rl97}. However, most of these sources, being of T\,Tauri type, are presumably at a 
later stage of development.

Source variability could be another, perhaps more likely, reason
and in which case neither of the two \mdot$_{\rm w}$ estimates would be entirely correct, since both
methods are based on the assumption of stationarity. Intensity variations from the shocked gas on small 
spatial and temporal scales (\lapprox\,10\asec, $<0.5$\,yr) have frequently been observed \citep{cameron90,rl96}
and the \oishort\ cooling time is likely of this order.
As discussed by \citet{giova00}, the H\,I data pertain to time scales of the order of 30\,yr and the memory of
small scale variations might become `ironed out' in the global flow. 

\subsection{The small scale flows} 

\subsubsection{The velocities of the optical jets}

On the sub-arcsec to ten arcsec scale, two independent jets from IRS\,5 have been identified in the 
optical and in the near infrared \citep{fl98,itoh00,mf04}. These have been designated as the `northern' 
and the `southern' jet, respectively. The two jets distinguish themselves by very different emissivity and velocity
characteristics.
The intensity of the northen jet is much stronger and the material exhibits much larger velocities.
At the projected jet position 2\asec\ (300\,AU) distant from IRS\,5, \citet{mf04} have recently obtained a 
maximum (blue-shifted) radial velocity of $\max v_{\rm rad} = v \cos{\theta} = 430$\,\kms. This
flow velocity has been maintained at this high level over the years \citep[cf.][]{stocke88}.
Radial velocity and proper motion data imply that the jet is moving at non-zero angles with respect 
to the line of sight, i.e. $\theta_{\rm N} > 0$ \citep{ls86,stocke88,fl94,lucas96},
so that the jet velocity is strictly higher than 430\,\kms, consistent with the shock generated X-ray luminosity
observed from this jet position \citep{fava02,bally03,mf04}. In contrast, the radial velocity observed for
the southern jet is at most 65\,\kms\ \citep{fl98,hart00,pyo02,mf04} and any non-zero proper motion is below
detectability in the multi-epoch data of \citet{mf04}.

\subsubsection{The densities of the northern optical jet}

At the position of observed highest radial velocity, \citet{mf04} have estimated local
volume densities from spectrophotometric observations of lines of [S\,II], viz. $n_{\rm e}
= (1.8 \pm 0.4) \times 10^3$\,\cmthree. This value represents the averaging over  
a region of effective radius 170\,AU.

For this region of the northern jet, \citet{itoh00} have obtained a spectrum in the wavelength interval 
1.05 to 1.82\,\um\ and identified eight [Fe\,II] lines.
These lines would be less affected by intervening extinction than the optical [S\,II] lines.
Also, as noted by \citet{itoh00}, most critical densities (ratio of radiative to collisional rates)
for [Fe\,II] are higher than those of [S\,II]. Consequently, the [Fe\,II] lines trace potentially 
gas at higher densities. 

\begin{figure}
\includegraphics[angle=00,scale=0.75]{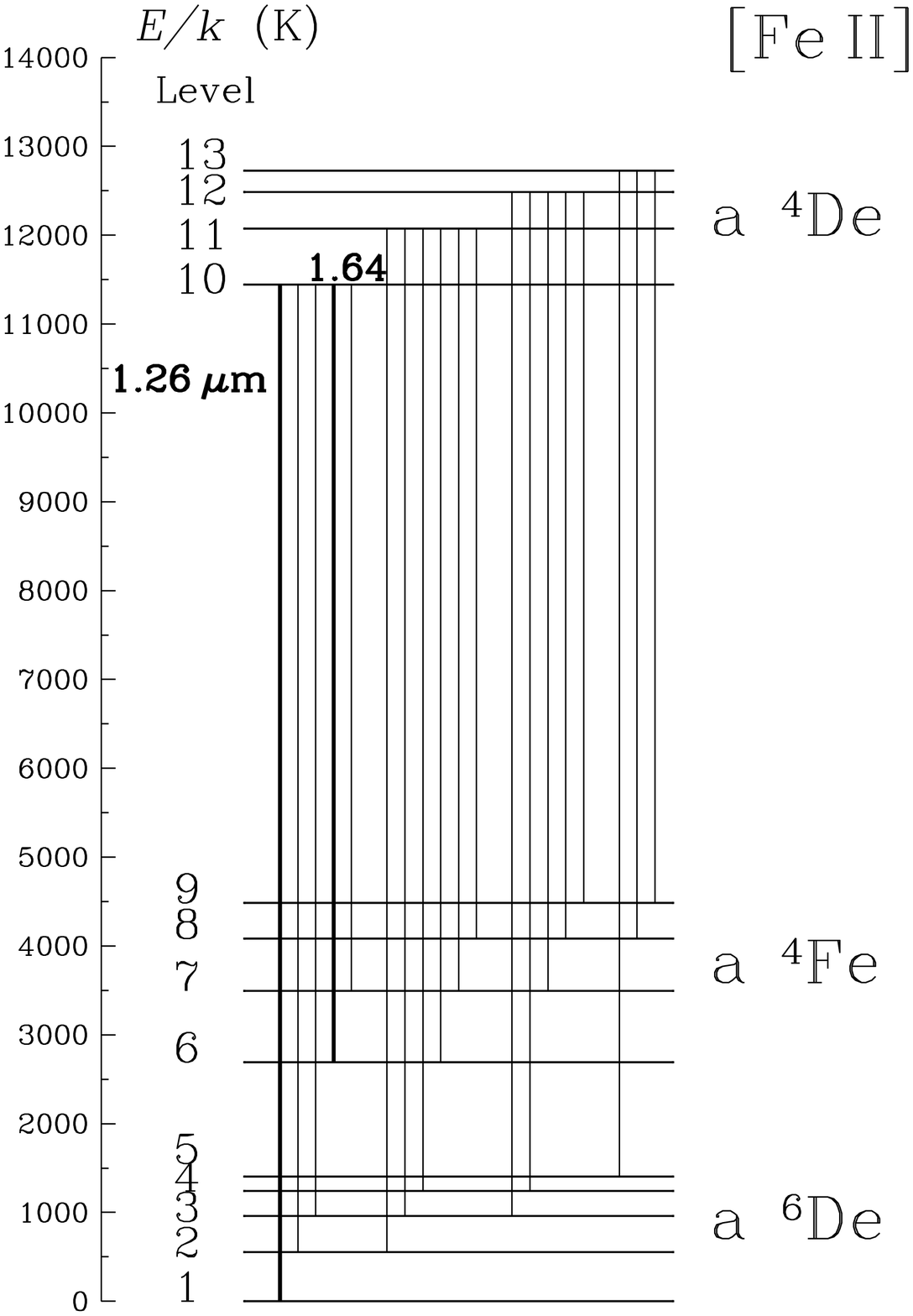}
\caption{The thirteen lowest energy levels of [Fe\,II] are displayed. The 19 transitions marked 
in the spectrum of Fig.\,\ref{Fe_spectrum} are shown by the lines connecting various levels. The two strongest
lines in that spectrum are shown by the thick lines.}
\label{energy_levels}
\end{figure}

We have extended the analysis by \citet{itoh00} by model fitting their entire observed spectrum, using 
an [Fe\,II] model atom having 142 energy levels with 1438 transitions and with atomic data from
\citet{quinet96} and \citet{zhang95}. The absolute flux calibration was kindly communicated to 
us by Y.\,Itoh. The transitions under consideration are identified in the energy level
diagramme of Fig.\,\ref{energy_levels}. 

\begin{figure}
\includegraphics[angle=270,scale=0.6]{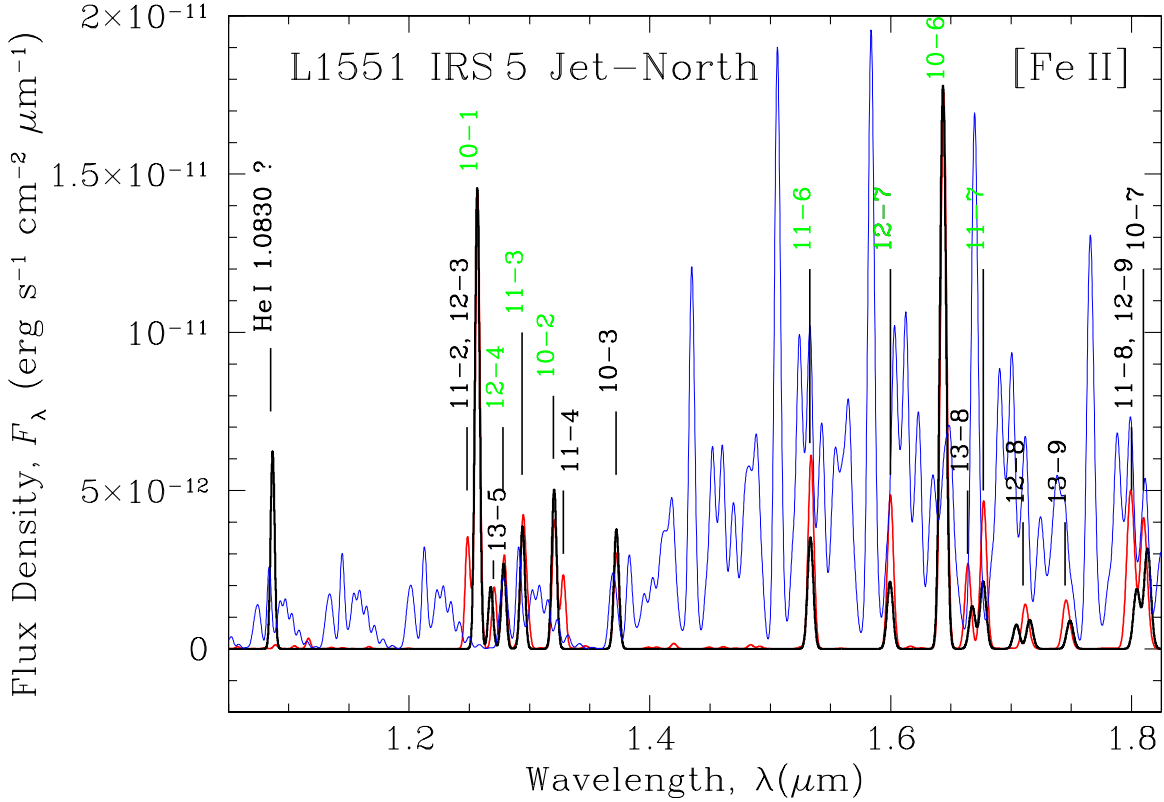}
\caption{The observed and synthetic spectra of the northern jet in the photometric J- and H-bands. 
The black curve corresponds to the flux densities calculated from the line flux observations of 
\citet[Table~1, ID=Eb, $R_{\lambda}=280$]{itoh00}, who identified 8 [Fe\,II] lines, the
transitions of which are labelled in green (cf. Fig.\,\ref{energy_levels}). Other line fluxes were estimated
from their Fig.\,5. 
The fitted model is shown in red: whereas the matching in the J-band is excellent, some of the weaker lines
in the H-band appear too strong in the model. This `mis-fit' could be due to insufficient
background correction. The OH lines are particularly intense in this wavelength region, as is illustrated 
by the blue spectrum, showing the arbitrarily scaled telluric OH emission \citep{rousselot00}.  
Since X-ray emission has been detected from this jet position \citep{fava02,bally03,mf04}, we tentatively 
identify the emission feature at 1.08\,\um\ as the recombination line ${\rm He\,I}\,\lambda 1.0830$ 
($^3{\rm P}^0\,2{\rm p}\, - \,^3{\rm S}\,2{\rm s}$).}
\label{Fe_spectrum}
\end{figure}

Our best fit model is shown in Fig.\,\ref{Fe_spectrum} for the spectral resolution of 280 \citep{itoh00}.
These calculations assume a Full Width Half Maximum of 150\,\kms\ for the [Fe\,II] lines \citep{pyo02,mf04}, solar 
chemical composition and an electron temperature $T_{\rm e}=10^4$\,K. Little is known about the
abundances in the jet, but the assumption of a solar iron abundance seems not too unreasonable
\citep{beck-win94,bohm01}. The absence of lines from highly excited states in the observed spectrum
limits the temperatures to significantly below $3 \times 10^4$\,K, a regime in which the final results
are not critically sensitive to the assumed temperature.

For optically thin emission (verified by the model calculations), the line ratio 
$I_{1.64\,\mu{\rm m}}/I_{1.25\,\mu{\rm m}} = {\rm constant} = 0.96$. For an average interstellar 
extinction law \citep{rieke85}, the observed line ratio of 1.6 implies a visual extinction, \av, of 
5.72 magnitudes. This value is consistent with the absence of detectable Paschen line emission,
based on the \halpha-flux of \citet{mf04}, $F_{\alpha}=4 \times 10^{-15}$\,\ecs, 
and on Case\,B recombination \citep{hummer87}. This \av-value is also an upper limit, since for
higher values, the P$\beta \,\lambda\,1.2818$\,\um\ line would be readily seen in the spectrum of 
\citet{itoh00}. Similar applies to P$\gamma \,\lambda\,1.0938$\,\um.

With the extinction fixed, the fitting of the observed spectrum results in a local jet density, 
formally, as $\log n_{\rm e} = 4.88 \pm 0.02$, i.e. $(7.6^{+0.3}_{-0.4})\times 10^4$\,\cmthree, 
for $1\sigma$ observational errors \citep{itoh00}. These densities in the northern jet are much 
higher than what can be determined for the southern jet.

The source size of this model is 60\,AU (jet diameter = \asecdot{0}{4}), the column density is 
$N_{\rm [Fe\,II]}=2 \times 10^{15}$\,\cmtwo\ and the total [Fe\,II] cooling rate is 
$L_{\rm [Fe\,II]}= 7 \times 10^{-4}$\,\lsun\ ($2.7 \times 10^{30}$\,\es).

\subsubsection{The radio jets}

On the \asecdot{0}{1} to arcsec scales, \citet{rod03a} have recently confirmed the binarity of the
L\,1551 IRS\,5 jet also at radio wavelengths (3.5\,cm). The radio jets appear well aligned
with their optical counterparts which become detectable only further `down-stream', because of heavy 
extinction. Given that the northern jet is the dominant one at most wavelengths,
it might seem enigmatic that the 3.5\,cm emission from the {\it southern} jet is stronger 
by a factor of about two. For free-free emission, a relatively mild,
by a factor of $\sqrt2$, variation of the electron density, integrated along the line of sight
($S_{\nu,\,{\rm free-free}} \propto \int (x_{\rm e} n_{\rm H})^2\,{\rm d}l$), could accomplish this. 
The southern jet does not conform with the generally adopted idea about the disk-jet geometry, 
as its direction appears to deviate substantially from the expected orthogonality \citep{rod03a}.

\subsection{Properties of the binary protostar}

On the basis of high resolution VLA observations at millimeter wavelengths, \citet{rod98} resolved
the central source IRS\,5 into two components. The marginally resolved emission peaks have an elliptical
appearance and most likely represent circumstellar disks around each of the components in 
a protostellar binary system. The protostars themselves are, of course, not directly detected
at the wavelength of 7\,mm.

\subsubsection{The photospheric spectrum}

The spectroscopic observations of the nebulosity HH\,102 (S\,239) by \citet{mundt85} 
revealed the reflected photospheric spectrum of the deeply embedded object IRS\,5. The 
wavelength range was extended by \citet{stocke88}, who concluded 
that the data were consistent with stellar spectral types of giants (luminosity class III)
ranging from G2 (in the blue spectral range) to K0 (in the green spectral region), with an uncertainty of two sub-classes. 
In addition, the absorption line depths were indicative of a surface gravity even lower than that of 
supergiant stars (luminosity class I). Taken to the extremes, this would allow for a considerable range 
in photospheric temperatures, viz. from 4300\,K (K2\,I) up to 5600\,K (G0\,III), see, e.g., \citet{cox99}. 
In our analysis below, we will adopt a slightly less conservative range, viz. 
$T_{\rm eff} = (5100 \pm 200)$\,K, which covers the spectral types K0\,III to G0\,III and which includes the
temperatures of supergiants of earlier spectral type. \citet{stocke88} interpreted this drift in spectral type,
in combination with the low gravity, as evidence for an FU Orionis (FUOR) type of disk around IRS\,5.

As already noted by \citet{mundt85}, and reinforced by \citet{stocke88}, the Balmer line profiles exhibit
P\,Cygni structure, indicative of a wind velocity of about 440\,\kms, essentially identical to 
the H$\alpha$ line width observed by \citet{mf04} in emission from the northern jet.

\subsubsection{The radiative luminosity}

For a distance of 150\,pc, detailed and self-consistent fitting of the entire observed spectral energy 
distribution (SED) of IRS\,5, using a two-dimensional radiative transfer model for a disk structure, led 
to the determination of the total luminosity, $L_{\rm tot}=40$\,\lsun\ \citep{white00}. This value is 
larger by nearly 40\% than the calorimetric luminosity, $L_{\rm cal}=30$\,\lsun, obtained from direct 
integration of the SED, owing to photon escape in the low-density polar directions. 
The model also correctly reproduces observed spatial intensity profiles and interferometric visibilities,
lending further confidence in the luminosity estimate by \citet{white00}.

\subsubsection{The dynamical mass of the system}

From multi-epoch radio observations of the proper motions of the binary source IRS\,5, 
\citet{rod03a} have estimated the dynamical mass of the system as
$0.1 \le \sum_{i} m_{\star,\,i}+ \sum_{i} m_{{\rm D},\,i} \sim 1.2$\,\msun.
The mass of the disks $\sum_{i} m_{{\rm D},\,i} = 0.1$\,\msun, with 
the northern disk being twice as massive as the southern one \citep{rod98}.

\section{Protostar models}

For observationally derived values of $\theta$ (45\adeg\ to 60\adeg, Sect.\,2.2.1), the velocity of the northern jet,
$> 600$\,\kms, is comparable to the escape velocities of main-sequence stars, given by

\begin{equation}
v_{\rm esc} = 617.7 \left ( \frac {M}{R}\right )^{1/2}\hspace*{0.2cm}M_{\odot}^{-1/2} \,R_{\odot}^{1/2}\,{\rm km\,s}^{-1}
\label{e1}
\end{equation}

with obvious notations. Since the combined mass of the binary amounts to about 1\,\msun, the total luminosity
should not exceed 1\,\lsun, if the stellar components were in the main-sequence. This is inconsistent
with observation ($L_{\rm tot}=40$\,\lsun) and attests to the pre-main-sequence nature of the objects.
One notes that this luminosity, on the one hand, is much too low for a typical FUOR, but also much too high for
a T\,Tauri star, on the other. 

It is illustrative to try to place IRS\,5 into the H-R diagram, as shown in Fig.\,7 of \citet{sst80}, where
it would be somewhere near the top of the curve labelled `gas photosphere'. This would also be consistent with
the age of the large scale molecular outflow, viz. $\sim 10^5$\,yr \citep{snell80,padman97}. In the figure by
\citet{sst80}, the evolutionary tracks for the protostellar `dust' and `gas photosphere', respectively, are
separated. However, unlike their case of isotropic emission, the protostellar photosphere of IRS\,5 can be 
viewed in reflection, since the dust `shell' is not optically thick in all directions. 

At the time \citet{stocke88} wrote their article, the binary nature of IRS\,5 was not established. As an
alternative to their suggestion, we shall below explore the possibility that the scattered light observed
by \citet{mundt85} and \citet{stocke88} is due to the combined spectrum of two protostellar photospheres.

\subsection{The mass-radius relations and accretion luminosities} 

At present, only the total luminosity, $L_{\rm tot}$, of the IRS\,5 system is determined observationally.
Assuming that the binary is protostellar in nature and, as such, derives its luminosity mostly from 
mass accretion processes, the total luminosity is the sum of the accretion luminosities
of each of the members of the system, i.e., with common notations,

\begin{equation}
L_{\rm tot} = \sum_{i} L_{{\rm acc},\,i} = 
               G \sum_{i} \dot{M}_{{\rm acc},\,i} \left ( \frac{M}{R} \right )_{i}\hspace*{0.2cm}. 
\label{e2}
\end{equation}

\begin{figure}
\includegraphics[angle=00,scale=0.75]{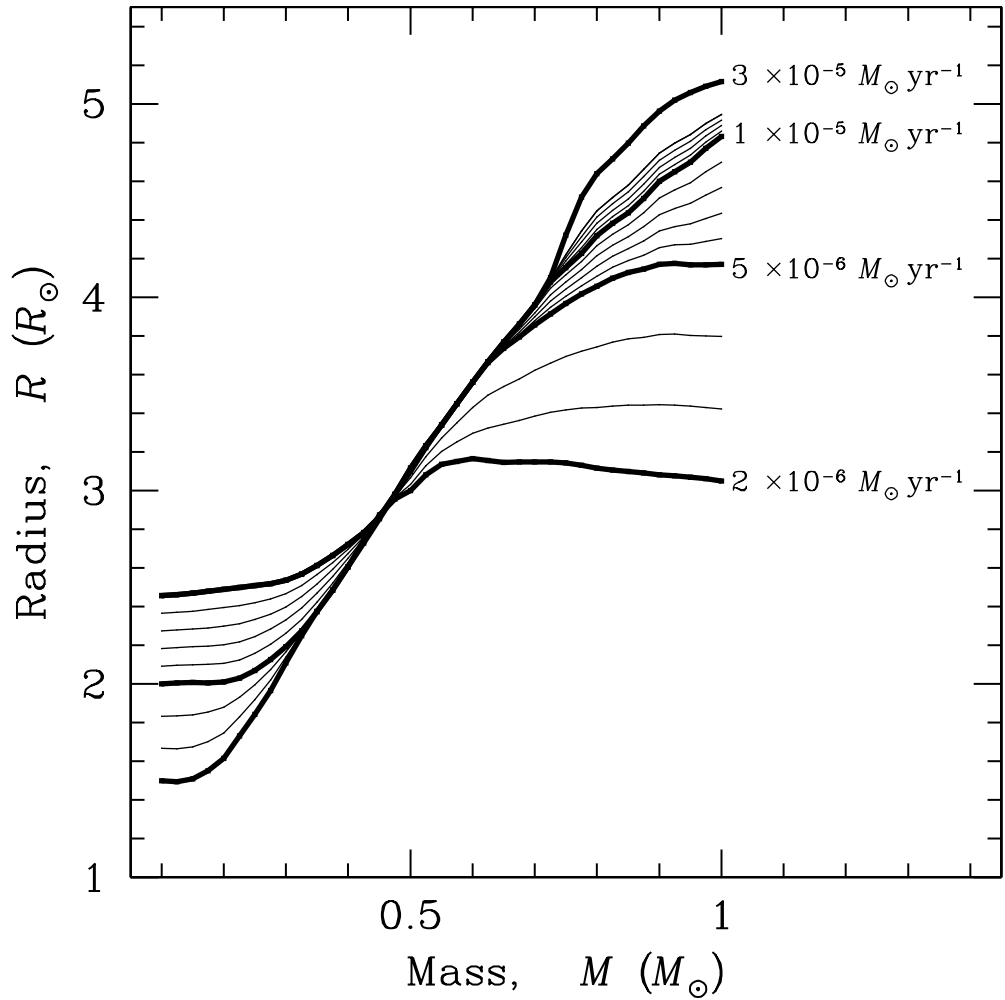}
\caption{Mass-radius relations of protostars during deuterium burning. The curves drawn with thick lines
are adopted from the literature, with our interpolations shown by the thin lines. The parameters next to
the curves are the mass accretion rates, with the data for $2 \times 10^{-6}$, $5 \times 10^{-6}$ and
$1 \times 10^{-5}$\,\msunyr\ taken from \citet{stahler88} and those for $3 \times 10^{-5}$\,\msunyr\ from 
\citet{ps92}.}
\label{mass-radius}
\end{figure}

We assume that the stellar contribution to $L_{\rm tot}=40$\,\lsun\ \citep{white00} is 
provided by two objects of total mass, $M_{\rm tot}= 1.1$\,\msun\ \citep{rod03b}, with a
minimum mass of about 0.1\,\msun\ for one of the components. Thus, for the
examination of Eq.\,\ref{e2}, we can limit the mass range for the individual masses 
to $M_i \in \left [ 0.1,\,1.0\right ]$\,\msun.
This is `fortunate', since in this mass interval the mass-radius relation, yielding the
ratio $(M/R)_i$ for a given mass accretion rate $\dot{M}_{{\rm acc},\,i}$, is insensitive to
the details of the accretion processes and physical boundary conditions 
\citep[see][and~Fig.~\ref{mass-radius}]{stahler88,ps92,ps93}, provided these objects are in
their deuterium burning stage.

The cited references provide mass-radius relations, $R(M)$, for a few mass accretion rates.
For other values, we obtained the mass-radius relations by interpolating in between
these rates (see Fig.\,\ref{mass-radius}).

\subsection{The surface luminosities and effective temperatures}

The effective temperatures of the protostellar photospheres are obtained from \citep{stahler88,ps93}

\begin{equation}
T_{{\rm eff},\,i} = \left ( \frac{L_{{\rm surf},\,i}}{4\pi\sigma R_i^2} \right )^{\frac{1}{4}}\hspace*{0.2cm}, 
\label{e3}
\end{equation}

where the surface luminosity, $L_{{\rm surf},\,i}$, is the sum of the radiative and convective contributions, given by

\begin{equation}
L_{{\rm surf},\,i} = 
L_{{\rm rad},\,i}+L_{{\rm conv},\,i} \sim L_0\,\sqrt{\frac{(M/M_{\odot})_i^{11}}{(R/R_{\odot})_i }} + 
L_{{\rm D},\,i}\hspace*{0.2cm},
\label{e4}
\end{equation}

with $L_0 = 0.153$\,\lsun, and where we have approximated $L_{{\rm conv},\,i}$ with the luminosity during full deuterium
burning, $L_{{\rm D},\,i}$. For accretion rates different from those given by \citet{stahler88}, 
$L_{{\rm D},\,i}$-values were obtained by interpolating the published $L_{\rm D}(M)$-curves. Since our primary
objective is to obtain some reasonable estimates of $T_{{\rm eff},\,i}$, we did not bother to attempt 
extrapolating the $L_{\rm D}(M)$-curves beyond those given by \citet{stahler88}. Instead,
for $1.1 \times 10^{-5}$ to $1.5 \times 10^{-5}$\,\msunyr\ we used the curve for $1 \times 10^{-5}$\,\msunyr, 
a procedure which will not affect our conclusions below. 

\subsection{Physical parameters of the binary components}

We examine numerically Eq.\,\ref{e2} on the intervals $M_i \in \,[0.1,\,1.0]$\,\msun\ and 
$\dot{M}_{{\rm acc},\,i} \in [2,\,15]\,10^{-6}$\,\msunyr, within which the parameters for both binary 
components are allowed to vary on the adopted grid, $\delta M=0.025$\,\msun\ and 
$\delta \dot{M}_{\rm acc}=1.0 \times 10^{-6}$\,\msunyr, respectively. This discretisation introduces
a `fuzziness' on the boundary condition $L_{\rm tot}=40$\,\lsun, which we estimate from
$(\Delta L/L)_{\rm acc} \sim \pm [(\partial M/M)^2+(\partial \dot{M}/\dot{M})^2 +(\partial R/R)^2]^{1/2}$ as about 
$\pm 15\%$ ($\pm 5.8$\,\lsun). This is comparable to the observational uncertainty. As an
additional constraint, the effective temperature of the {\it more luminous} component is bounded by the interval 
$T_{{\rm eff},\,i} \in \,[4900,\,5300]$\,K (see Sect.\,2.3.1).

\begin{figure}
\includegraphics[angle=90,scale=0.75]{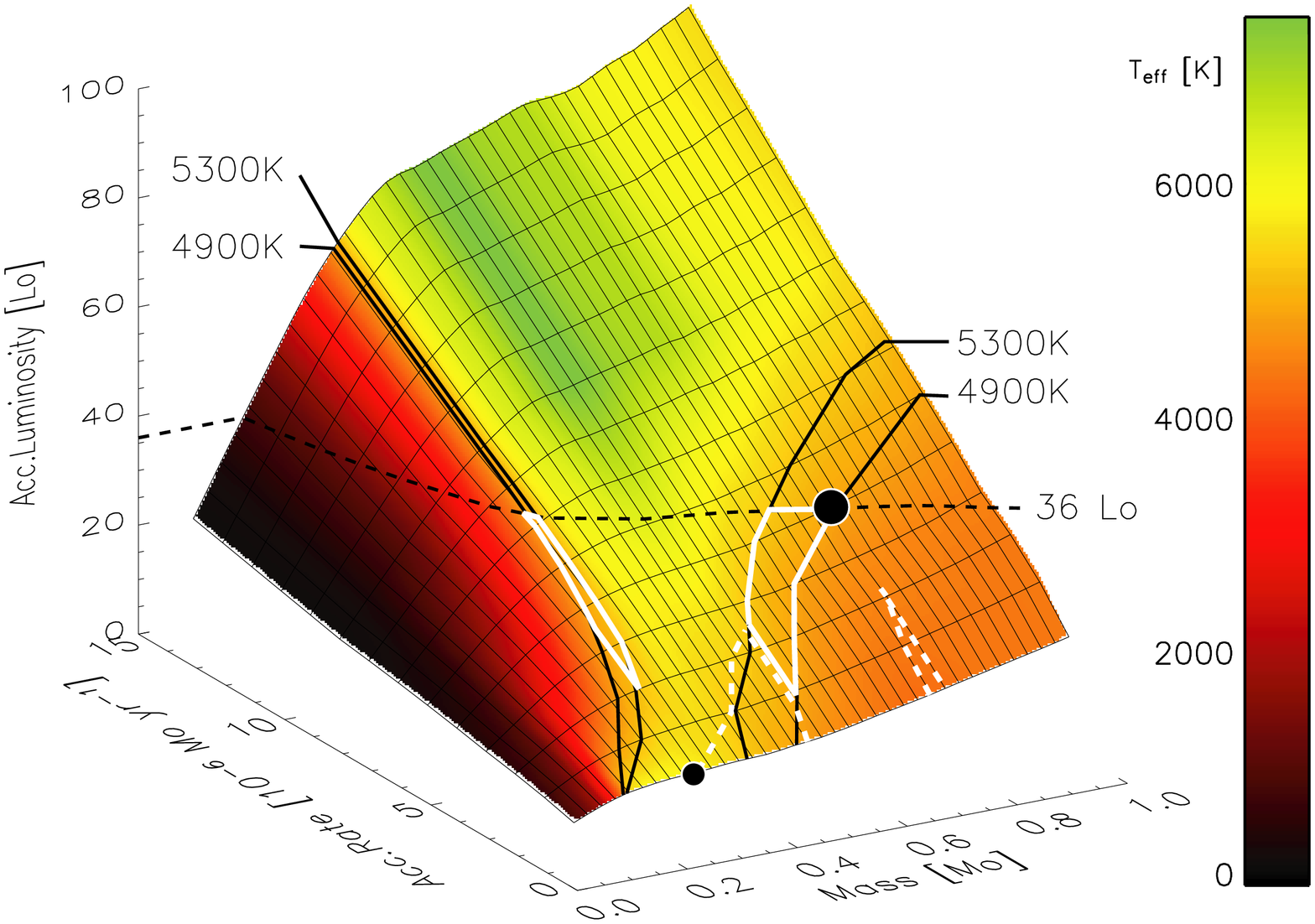}
\caption{The explored parameter space in the variables mass accretion rate, accretion luminosity, effective
temperature and  protostellar mass. Here, the radii of the protostar models are implicit and presented 
separately in Fig.\,\ref{mass-radius}. The bar to the right of the graph provides the colour coding
for the effective temperature. Areas, thought best to represent the protobinary L\,1551 IRS\,5 are encompassed
by the white lines, where the full-drawn lines indicate the positions of the primaries and the dashed lines 
those of the companions. The smaller and clearly separated areas correspond to pairs in which the less massive 
star is the more luminous object.
The black lines are meant to facilitate the viewer's orientation along the axes. Our best solution for the
IRS\,5 binary is shown by the fat black dots, with the bigger one identifying the primary (IRS\,5-N, see the text).}
\label{4D_fig}

\end{figure}

We find a couple of islands of formally acceptable solutions, shown in Fig.\,\ref{4D_fig}. There, the explored 
parameter space for the variables mass, accretion rate, accretion luminosity and effective temperature is depicted.
Most solutions select the more massive star (the `primary') as also the more luminous one. This is shown
by the larger area encompassed by the full-drawn white curve. The adjacent dashed lines outline the region, where
the corresponding secondaries are situated.

A smaller number of other solutions were also found, where the less massive star (the `secondary') is the more 
luminous one, because of a much larger mass accretion rate (and, hence, becomes the primary in the commonly accepted
sense). These are shown by the two separated islands.

Our preferred solution is shown in Fig.\,\ref{4D_fig} by the two black dots, where the larger one signifies
the primary (both more massive and more luminous). The reasons for this selection will become apparent below.
The physical parameters for the protostellar primary of IRS\,5 are: 
$M_1 = 0.8$\,\msun, 
$\dot{M}_{{\rm acc,}\,1} = 6 \times 10^{-6}$\,\msunyr, 
$T_{\rm eff,\,1} = 4.9 \times 10^3$\,K,
$\log g_1 = 3.1$ and
$v_{\rm esc,\,1} = 270$\,\kms.
For the secondary, the corresponding values are $M_2 = 0.3$\,\msun, 
$\dot{M}_{{\rm acc,}\,2} = 2 \times 10^{-6}$\,\msunyr, 
$T_{\rm eff,\,2} = 5.8 \times 10^3$\,K,
$\log g_2 = 3.2$ and
$v_{\rm esc,\,2} = 235$\,\kms, respectively.
The secondary contributes 25\% to the total luminosity and the
derived values of $\log g$ correspond to those of normal giants (luminosity class III, cf. Sect.\,2.3.1).

\subsubsection{Photospheric emission}

The parameters derived in the previous section can be used to predict the photospheric spectra of IRS\,5.
Simple estimates of the resulting monochromatic luminosities indicate that the photospheric emission 
of the hotter secondary should
be comparable in intensity to that of the cooler primary in the blue spectral region, but weaker already in the
green (and for longer wavelengths). This is verified in detail when examining theoretical stellar atmosphere models. 
We use the NextGen models of \citet{hauschildt99} for the appropriate effective temperatures, the closest available
values of the surface gravity ($\log g=3.5$) and solar chemical composition. 

\begin{figure}
\includegraphics[angle=90,scale=0.75]{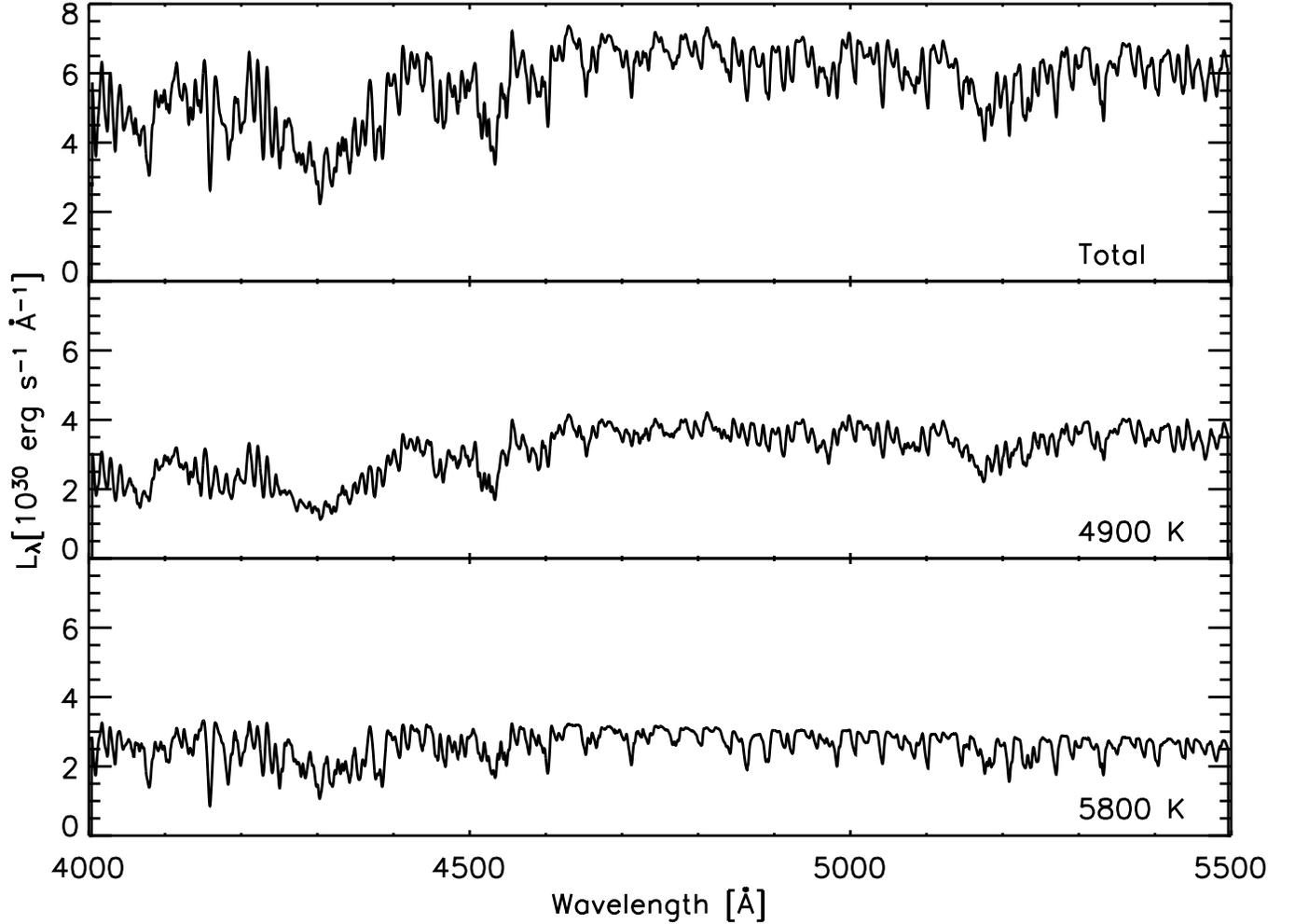}
\caption{Stellar model atmospheres from \citet{hauschildt99} for $\log g=3.5$, solar composition and the 
indicated $T_{\rm eff}$, representing (approximately) the preferred binary solution shown in Fig.\,\ref{4D_fig}.
The hotter secondary is comparable in brightness to the cooler primary in the blue spectral regime, resulting
in the relatively shallower G-band near 4300\,\AA\ than what would be expected on the basis of the green spectrum,
where the primary is dominating the photospheric emission. The upper curve is the sum of the spectra in the 
two lower panels.}
\label{spec}
\end{figure}

\begin{figure}
\includegraphics[angle=90,scale=0.75]{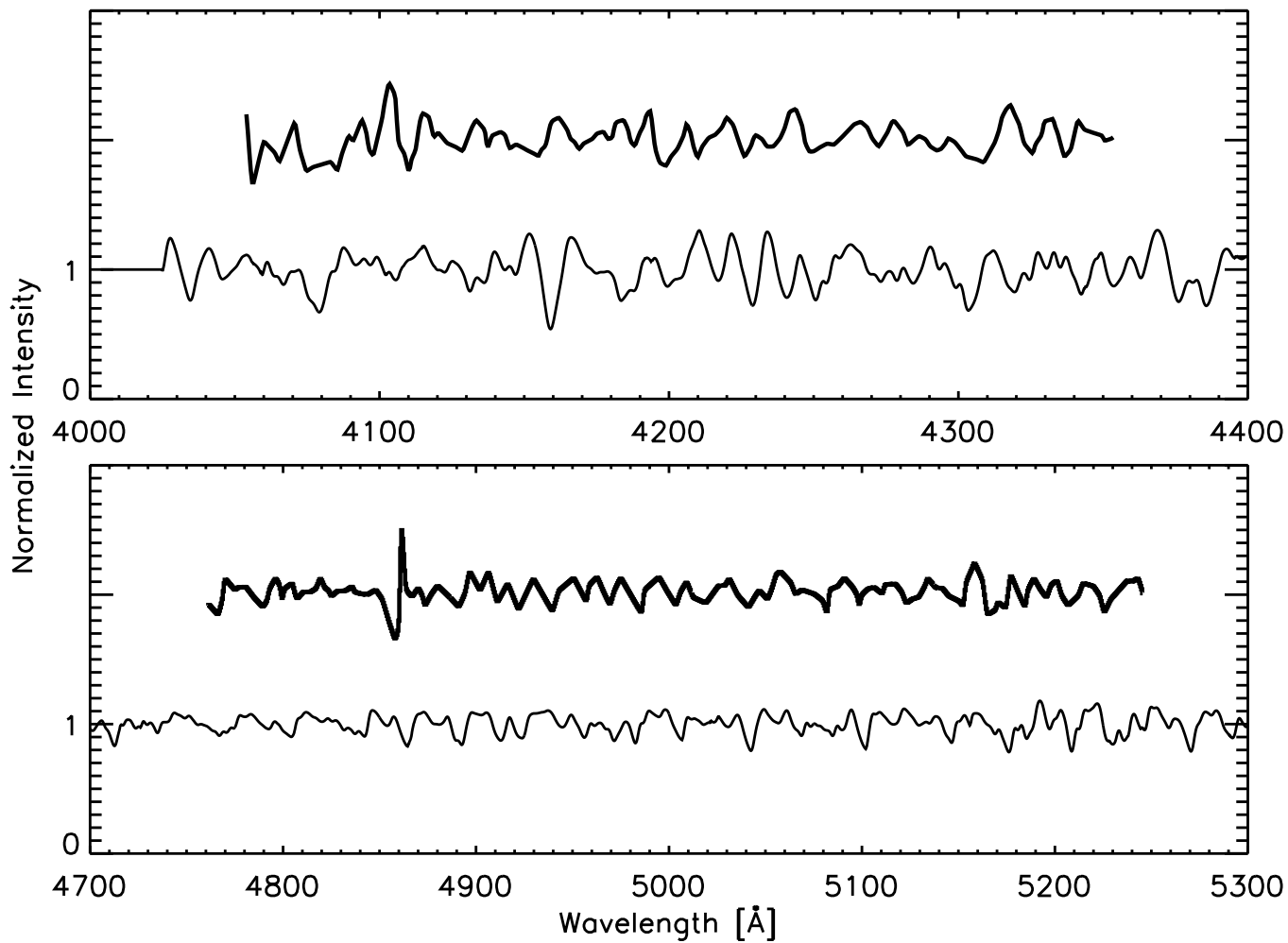}
\caption{The normalised observed optical spectrum of IRS\,5 \citep{stocke88} shown by the thick lines
and the composite binary spectrum (thin lines). The latter is based on the models shown in Fig.\,\ref{spec}, but
with their spectra rotationally broadened according to their escape velocities, prior to summation (see the text). 
Several features, such as H$\delta\,4101$ and H$\beta\,4861$ in particular, seem not to match very well, which can be 
ascribed to contaminating emission from the observed nebula HH\,102 \citep{stocke88}.}
\label{norm_spec}
\end{figure}

The individual model spectra and their sum are shown in Fig.\,\ref{spec}, whereas in Fig.\,\ref{norm_spec},
the normalized composite model spectrum is displayed together with the observations by \citet{stocke88}. The latter
models have been `spun up' to the escape velocities of the individual components prior to adding them together.
These velocities must be regarded as upper limits to real and observable rotation speeds (break-up and viewing geometry,
respectively), but these values are consistent with the limited spectral resolution of the observations of 
\citet{stocke88}. 

Rotation at, or rather close to, break-up in the stellar equatorial regions would also naturally explain the extreme
supergiant characteristics of some of the absorption lines, since in this case, $\log g$ corresponds by definition to
zero gravity.

For the interpretation of the nature of IRS\,5, the implications of the FUOR-disk and protostellar photosphere 
scenarios are very different. High resolution spectroscopy in the optical should enable us to distinguish between 
these alternate models and, furthermore, potentially provide the opportunity to study stellar surfaces during an 
evolutionary phase which has not previously been accessible to direct observation.

\subsection{Outflow models}

To put the observed and derived properties of the binary protostar and the binary jet
into context, we will make use of theoretical models of outflows. Specifically, the theory
of magnetocentrifugally driven flows (x-winds) has been worked out in considerable detail and
presented by F.\,Shu and co-workers in a series of papers.

\subsubsection{x-wind velocities}

In terms of the mass-radius relation (Sect.\,3.1), the expression for the terminal wind velocity 
by \citet[their Eq.\,4.13a]{shu94} can be recast into

\begin{equation}
{\overline v}_{\rm w} \le \left [ \,\frac {2\,{\overline J} -3}{r_x} 
\,\,G \,\left (\frac{M}{R} \right )\right ]^{1/2}\hspace*{0.2cm}, 
\label{e5}
\end{equation}

where $r_x$ is a factor, such that $r_x$ times the stellar radius $R$ is the
radial distance of the location of the x-point, assumed to be close to the
radius of corotation of the stellar surface and the circumstellar Keplerian disk,
so that $r_x$ is of order unity. The star is assumed to rotate near break-up. 

\begin{figure}
\includegraphics[angle=00,scale=0.75]{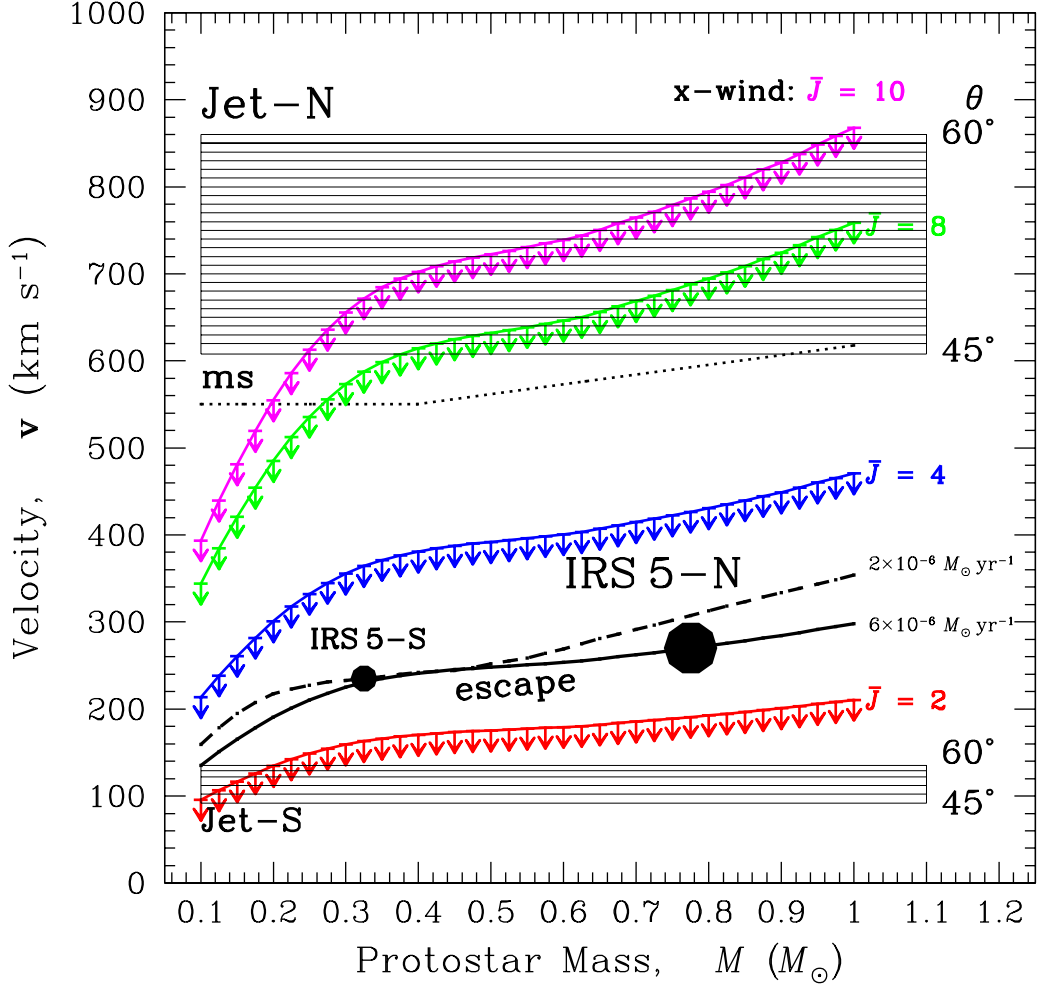}
\caption{Various velocities versus protostellar mass. The deprojected radial velocities of the jets 
from L\,1551 IRS\,5 are shown by the shaded regions labelled Jet-N and Jet-S, respectively. The dots
denote the positions of the corresponding driving sources, viz. the primary IRS\,5-N and its less
massive companion IRS\,5-S. These are shown on curves of the escape velocity, corresponding
to their mass accretion rates of $6 \times 10^{-6}$ and $2 \times 10^{-6}$\,\msunyr, respectively.
For comparison, the escape velocities for the main-sequence (ms) are also shown by 
the dotted line. The curves with the upper limit symbols are the loci of the asymptotic x-wind
velocity for different values of the angular momentum parameter $\overline{J}$ and the accretion rate of
$6 \times 10^{-6}$\,\msunyr. To reach velocities as high as those of Jet-N, large values of $\overline{J}$ 
are required, implying $f$-values of 0.3 or smaller (see the text).}
\label{v_escape}
\end{figure}

As explicitly indicated, Eq.\,\ref{e5} represents an 
upper limit to the wind velocity, with equality reached at infinity \citep{shu95}. 
The angular momentum parameter, ${\overline J}$, takes values $>1.5$ 
and is likely not to exceed $\sim 10$. 
For the mass-radius relation corresponding to the accretion rate of $6 \times 10^{-6}$\,\msunyr\
(Fig.\,\ref{mass-radius}), graphs of Eq.\,\ref{e5}, with $r_x=1.0$, are shown in Fig.\,\ref{v_escape}.

\subsubsection{x-wind $f$-factors and densities}

For a single stellar mass and a particular choice of parameters ($M=0.5$\,\msun, $R=4$\,\rsun, 
$r_x=1.0$, \mdot$_{\rm w}=10^{-6}$\,\msunyr), \citet{ns94} presented detailed numerical results, 
covering the range in $\overline J$ from 2.0 to 7.8, with accompanying magnetic field strengths from 1.5 to 8.3\,kG.
Corresponding $f$-factors vary from 0.4 to 0.1, where $f$ is defined as the ratio of the mass loss to the
mass accretion rate.

The collimation of, initially wide-angled, x-winds into narrow jets has been addressed by \citet{shu95}.
These authors also make a theoretical prediction of the density profile across the jet, i.e. in terms of the 
jet radius, the density scales approximately as $\rho(r) \propto r^{-2}$. 

\section{Discussion} 

\subsection{The deuterium abundance}

The results presented for the protostar binary IRS\,5 are based on the mass-radius relations during 
deuterium burning as provided by the models of \citet{stahler88} and \citet{ps92}. These were calculated 
for the deuterium abundance relative to hydrogen, ${\rm [D/H]} = 2.5 \times 10^{-5}$, and are, as such,
sensitive to the assumed [D/H]. 

Recent observations indicated the significantly lower value of about
$1.5 \times 10^{-5}$ in the solar neighbourhood and, furthermore, with a considerable dispersion 
\citep{sonneborn02,moos02,steigman03}. However, \citet{lw04} recently announced that they could reconcile
the observational data with a value of ${\rm [D/H]} = (2.3 \pm 0.4) \times 10^{-5}$ within
1\,kpc of the Sun, in which case the results of the calculations by \citet{stahler88} and \citet{ps92},
with regard to the assumed deuterium abundance, remain valid.

\subsection{Single-star theory and binary system}

We obtain a binary mass ratio of $q = M_2/M_1 \sim 0.4$, which is close to the peak of the observed distribution 
for solar-type stars in the solar neighbourhood \citep{dm91}. These authors did also not find any
statistically significant dependence of the $q$-distribution on binary period (hence, binary separation).

Both the photospheric spectrum and the jet momenta are not easily reconcilable with an equal-mass binary,
accreting at the same rate. Also the higher disk mass of the primary would heuristically be 
compatible with the higher $\dot{M}_{{\rm acc},\,1}$ 
%($\dot{M}_{\rm acc} \propto \Sigma_{\alpha-\rm disk}$,
\citep[$\dot{M}_{\rm acc} \propto \Sigma_{\alpha-\rm disk}$,~e.g.][]{frank85}, 
since both disks have the same size \citep[about~10~AU,~][]{rod98}. 

Above, we have tacitly assumed that the components of the protostellar binary can be modelled with the
theory of single objects. The relatively large separation \citep[$43\,{\rm AU}/\cos \theta_{\rm binary} 
> a_{\rm crit}$;~][]{rod03a,hale94} justifies this assumption. Here, $a_{\rm crit} \sim 30-40$\,AU 
is the `critical separation', beyond which observed solar-type binaries cease to be coplanar \citep{hale94}, 
and hence can be assumed to have developed independently.

\subsection{The binary jet-driving IRS\,5 and x-wind models}

Each component of the protostellar binary drives its own jet. The (total) wind mass loss rates 
of Sect.\,2.1 are therefore likely upper limits to the individual loss rates.
Comparing with the result of Sect.\,3.3, the ratio of the mass loss to accretion rate for the primary
is estimated at $f_1 = (\dot{M}_{\rm w}/\dot{M}_{\rm acc})_1 < 0.13$ ($< 0.33$), where the value in 
parentheses refers to the H\,I rate. Correspondingly, for the secondary, $f_2 < 0.4$ ($< 1.0$).

When compared to x-wind models, these values are consistent with high $\overline J$ ($\sim 8 - 10$)
and low $\overline J$ ($\sim 2$), respectively \citep[see their Table~5]{ns94}. The inclination corrected
jet velocities conform with this scenario (see Fig.\,\ref{v_escape}). Based on our analysis, 
we identify the driver of the fast and well-aligned northern jet with the protostellar primary 
(IRS\,5-N) and, consequently, the source of the much slower southern jet with the secondary (IRS\,5-S).

Unless the disks are occasionally rejuvenated by the significantly more massive 
circumbinary reservoir \citep[and~references~therein]{osorio03,mf02}, 
the current level of the mass loss through the wind could be maintained for at most another 
$M_{{\rm D,}\,1}/\dot{M}_{{\rm acc,}\,1} \sim 0.06\,M_{\odot}/6 \times 10^{-6}\,M_{\odot}\,{\rm yr}^{-1} = 10^4$\,yr. 
This would be only about one tenth of the age of the large scale CO outflow, but comparable with the time scale 
(\about\,\powten{3}\,yr) of the molecular outflow close to the source \citep{fk93}. 
Given the large difference in their momentum rates, these two flow phenomena appear only indirectly related. 
The scenario which emerges \citep{mf04} is that the 
molecular flow lifts off the disk at some AU distance \citep[disk-wind,~e.g.,][]{kp00}, whereas the atomic 
flows/jets originate much closer to the protostellar surface (x-wind).

The results of Sect.\,2.2.2 can be used to examine the density behaviour {\it across} the northern jet. The
observed difference in density on different spatial scales hints at a non-flat density distribution in the
jet. These estimates of the density are based on optically thin line emission and, as
such, represent averages over volume, i.e. $\overline {n} = \int\!n\,{\rm d}V/\int\!{\rm d}V$. We assume that 
the density distribution perpendicular to the jet axis can be approximated by a power law, $n(r) \propto r^{\alpha}$.
If we examine a sufficiently small section of the cylindrical jet far away from the source, so that 
the density distribution perpendicular to the jet-axis does not depend on the height-coordinate ($z$-coordinate)
of the cylinder, the integrals become trivial and
$\overline {n} = 
2 n(r_0) (R/r_0)^{\alpha} \left [ 1 - \left ( r_0/R \right )^{\alpha + 2}\right ]/(\alpha + 2)$, $\alpha \ne -2$ ,
and $\overline {n} = 2 n(r_0) (R/r_0)^{\alpha} \ln {\left ( R /r_0\right )}$, $\alpha = -2$ ,
where $r_0 \ll R$ is some fiducial radius close to the central jet axis. This recovers the well known result that the
average density is dominated by the largest scales. In either of these two cases, the power law exponent is given
approximately by $\alpha \sim \Delta \log{ \overline {n}}/\Delta \log{R}$, as usual. Our line data
represent post-shock values for gas which has been compressed by an, in general, unknown amount. However, 
for a planar cross section, the compression is constant over the shock surface and from the [Fe\,II] and [S\,II] 
data we then obtain formally $\alpha = -2.3^{+0.1}_{-0.2}$ for the observed, and resolved \citep{mf04}, scales 
$R_{\rm [Fe\,II]}$\,=\,\asecdot{0}{2} and $R_{\rm [S\,II]}$\,=\,\asecdot{1}{0}. This observationally determined
value of the power law exponent is in reasonable agreement with the theoretical one for x-winds, viz.
$\alpha \rightarrow -2$ as $z \rightarrow \infty$ \citep{shu95}.

This circumstance motivates the direct comparison with the predictions by the x-wind model. With the
formalism of \citet{shu95} and for the parameters of IRS\,5-N, we estimate that the central density (at $r_0 = R_x$)
is $\rho_x \le 1.8 \times 10^{-12}$ ($4.6 \times 10^{-12}$) g\,${\rm cm}^{-3}$,
with the same convention regarding the mass loss rate as before. At the deprojected distance from the 
protostar of about $400-600$\,AU and at the radial distance from the jet axis $r=30$\,AU, 
we estimate from Fig.\,3 of \citet{shu95} that the jet density should be
of the order of $6 \times 10^{-20}$ ($15 \times 10^{-20}$) g\,${\rm cm}^{-3}$, which translates into a 
neutral particle density of $4 \times 10^4$ ($9 \times 10^4$) \cmthree\ of unshocked gas. 
As already remarked above, the observationally derived value ($\sim 7.5 \times 10^4$\,\cmthree) refers to 
the post-shock electron density, assuming an ionisation fraction of unity. 
The comparison of these two values suggests that the compression of the shocked gas
is just offset by the fractional ionisation of the neutral jet flow. This seems reasonable, as
the former is, on the average, a factor of about 80 \citep{hollen89}, whereas the latter is a few percent \citep{shang04}.

\citet{shang04} have recently applied the x-wind model to the radio jet emission from IRS\,5. These authors 
focus on the southern jet because of its apparent higher radio flux density. However, this circumstance 
is not reflected at other wavelengths, including X-rays, optical, infrared, millimeter and 
the radio regime \citep[for~the~latter,~see][]{rod98}, at which the northern source and jet dominate. 
In addition, the flux difference is
small enough that it could be easily absorbed in the model by \citet{shang04} and we conclude that the x-wind 
theory provides a viable model which is capable of explaining various independent pieces of observational evidence.
A final and more decisive test of the theory would include the direct measurement of the magnetic field and the 
protostellar rotation rate.

\section{Conclusions}

Our main conclusions from this work can briefly be summarised as follows:

\begin{itemize}
\item[$\bullet$] Interpreting recent observational achievements within the framework of the theory of
protostellar structure and evolution allows the derivation of the physical properties of the individual
components of the binary protostar L\,1551 IRS\,5.
\item[$\bullet$] We offer an alternate scenario to the FUOR-disk hypothesis to explain optical 
spectroscopic data of the scattered light from IRS\,5, which can be put to test with currently available 
observing facilities. This offers potentially the unique opportunity to directly observe the surfaces of
accreting protostars.
\item[$\bullet$] The derived properties of the binary, in combination with their observed mass loss rate,
lead to results which are consistent with detailed predictions by the theory of x-winds.
\end{itemize}

\acknowledgements
We thank the anonymous referee for her/his thoughtful comments. RL enjoyed interesting
discussions with Mary Barsony, G\"oran Olofsson and Frank Shu. We are grateful to Francesco Palla
for making available to us the mass-radius relations for different shock boundary conditions.

\clearpage

\clearpage

\clearpage

\clearpage
\clearpage

\clearpage

\clearpage

\end{document}